\documentclass[aps,prx,twocolumn,showpacs,superscriptaddress,amsmath,amssymb,amsfonts,floatfix,longbibliography]{revtex4-1}

\usepackage{graphicx}
\usepackage{placeins}
\usepackage{float}
\usepackage{dcolumn}
\usepackage{bm}
\usepackage{amssymb}
\usepackage[version=4]{mhchem}
\usepackage{microtype}
\newcolumntype{C}[1]{>{\centering\arraybackslash}p{#1}}
\usepackage{xfrac}
\usepackage{gensymb}
\usepackage{xcolor}
\usepackage{multirow}
\usepackage{comment}
\usepackage{times}
\usepackage[varg]{txfonts}
\usepackage{mathrsfs}
\usepackage{amsmath,amssymb,bm,mathtools}
\usepackage{amsfonts}
\usepackage{booktabs}
\usepackage{natbib,hyperref}
\hypersetup{
	colorlinks=true, 
	linkcolor=blue,  
}
\usepackage{soul}
\usepackage{epstopdf}
\usepackage{array}
\usepackage{multirow}
\usepackage{tcolorbox}
\usepackage[percent]{overpic}
\usepackage{bbm}

\usepackage{widetable}
\usepackage{array}
\usepackage{xcolor,colortbl}
\usepackage{lipsum}

\begin{document}

\title{Parallel Spin Stripes and Their Coexistance with Superconducting Ground States \\ at Optimal and High Doping in La$_{1.6-x}$Nd$_{0.4}$Sr$_x$CuO$_4$}

\author{Qianli Ma}
\affiliation{Department of Physics and Astronomy, McMaster University, Hamilton, Ontario, L8S 4M1, Canada}
\email[Email: ]{maq1@mcmaster.ca}

\author{Kirrily~C. Rule}
\affiliation{Australian Nuclear Science and Technology Organisation,
Locked Bag 2001, Kirrawee DC, NSW, 2232, Australia}

\author{Zachary~W. Cronkwright}
\affiliation{Department of Physics and Astronomy, McMaster University, Hamilton, Ontario, L8S 4M1, Canada}

\author{Mirela Dragomir}
\affiliation{Department of Chemistry and Chemical Biology, McMaster University, Hamilton, Ontario, L8S 4M1, Canada}
\affiliation{Brockhouse Institute for Materials Research, Hamilton, Ontario, L8S 4M1, Canada}

\author{Gabrielle Mitchell}
\affiliation{Department of Physics and Astronomy, McMaster University, Hamilton, Ontario, L8S 4M1, Canada}

\author{Evan M. Smith}
\affiliation{Department of Physics and Astronomy, McMaster University, Hamilton, Ontario, L8S 4M1, Canada}

\author{Songxue Chi}
\affiliation{Neutron Scattering Division, Oak Ridge National Laboratory, Oak Ridge, TN 37830, United States}

\author{Alexander~I.~Kolesnikov}
\affiliation{Neutron Scattering Division, Oak Ridge National Laboratory, Oak Ridge, TN 37830, United States}

\author{Matthew B. Stone}
\affiliation{Neutron Scattering Division, Oak Ridge National Laboratory, Oak Ridge, TN 37830, United States}

\author{Bruce~D.~Gaulin}
\affiliation{Department of Physics and Astronomy, McMaster University, Hamilton, Ontario, L8S 4M1, Canada}
\affiliation{Brockhouse Institute for Materials Research, Hamilton, Ontario, L8S 4M1, Canada}
\affiliation{Canadian Institute for Advanced Research, MaRS Centre, West Tower 661 University Ave., Suite 505, Toronto, ON, M5G 1M1, Canada}

\date{\today}


\begin{abstract}
Quasi-two dimensional quantum magnetism is clearly highly correlated with superconducting ground states in cuprate-based High \textit{T}$_c$ superconductivity. Three dimensional, commensurate long range magnetic order in La$_2$CuO$_4$ quickly evolves to quasi-two dimensional, incommensurate correlations on doping with mobile holes, and superconducting ground states follow for \textit{x} as small as 0.05 in the La$_{2-x}$Sr$_x$/Ba$_x$CuO$_4$ family of superconductors. It has long been known that the onset of superconducting ground states in these systems is coincident with a remarkable rotation of the incommensurate spin order from ``diagonal stripes" below \textit{x} = 0.05, to ``parallel stripes" above.  However, little is known about the spin correlations at optimal and high doping levels, where the dome of superconductivity draws to an end.  Here we present new elastic and inelastic neutron scattering measurements on single crystals of La$_{1.6-x}$Nd$_{0.4}$Sr$_x$CuO$_4$ with \textit{x} = 0.125, 0.19, 0.24 and 0.26, and show that two dimensional, quasi-static, parallel spin stripes are observed to onset at temperatures such that the parallel spin stripe phase envelopes all superconducting ground states in this system.  Parallel spin stripes stretch across 0.05 $<$ \textit{x} $<$ 0.26, with rapidly decreasing moment size and onset temperatures for \textit{x} $>$ 0.125.  We also show that the low energy, parallel spin stripe fluctuations for optimally doped \textit{x} = 0.19 display dynamic spectral weight which grows with decreasing temperature and saturates below its superconducting \textit{T}$_c$.  The elastic order parameter for \textit{x} = 0.19 also shows plateau behavior coincident with the onset of superconductivity.  This set of observations assert the foundational role played by two dimensional parallel spin stripe order and fluctuations in High \textit{T}$_c$ cuprate superconductivity.

\end{abstract}

\maketitle

\section{Introduction}
The microscopic mechanism underlying high temperature superconductivity in copper oxide materials has been hotly debated since its original discovery in 1986 \cite{discovery1986,bkg2lbco}, through to the present day \cite{bkg1,theoryofhightc,bourne1987complete,kotliar1988superexchange,schrieffer1988spin,zhang1988effective,franck1994experimental,emery,kulic2000interplay,patricklee,mukuda2011high,scalapino,taillefer2010scattering,proust2019remarkable}. The magnetism associated with an approximately two dimensional square lattice of CuO$_2$ has figured prominently in this debate, both as a mechanism for superconducting pairing between electrons \cite{emery,patricklee,scalapino,keimer2015quantum}, and as a possible competing ground state \cite{corboz2014competing,rotatedstripe}.

Hole-doped High \textit{T}$_c$ cuprates, such as La$_{2-x}$Sr$_x$CuO$_4$ (LSCO), La$_{2-x}$Ba$_x$CuO$_4$ (LBCO), and La$_{1.6-x}$Nd$_{0.4}$Sr$_x$CuO$_4$ (Nd-LSCO), are in some ways the ultimate quantum materials \cite{quantummaterial}.  They display bulk superconductivity, and their parent compound, La$_{2}$CuO$_4$, is a stacked, two dimensional quantum magnetic insulator, based on \textit{S} = 1/2 Cu$^{2+}$ spin degrees of freedom.  La$_{2}$CuO$_4$ orders near 300 K into a three dimensional (3D) commensurate (C) antiferromagnetic (AF) structure \cite{Kastner,keimer1992magnetic}.  However, this 3D C AF order is very unstable to the presence of mobile holes, as are introduced by substitution of either Sr$^{2+}$ for La$^{3+}$ in LSCO and Nd-LSCO, and by substitution of Ba$^{2+}$ for La$^{3+}$ in LBCO.  3D C AF order is destroyed by \textit{x} $\sim$ 0.02, replaced by quasi-two dimensional (2D) incommensurate (IC) order with much lower onset temperatures \cite{Kastner,keimer1992magnetic,Kastner,3dAF2,3dAF3,3dAF4}.

At low dopings, for \textit{x} $<$ 0.05, the quasi 2D IC structure corresponds to ``diagonal stripes" wherein local regions of $\pi - \pi$ antiferromagnetism are partitioned into domains which are finite along (1,1,0)$_{Tetragonal}$ directions, that is at 45$\degree$ to the Cu-O bonds within the 2D plane \cite{keimer1992magnetic,wakimotostripe}.  These domains are separated by domain walls running along these diagonal directions, where the excess holes reside.  The domain walls introduce a $\pi$ phase shift into the AF structure, thereby giving rise to IC antiferromagnetism.  While such magnetic structures do not display true long range order, they do possess in-plane correlation lengths over 10\textit{s} of unit cells, and are quasi-static, such that Bragg peak-like features, well defined in \textbf{\textit{Q}} and elastic on the time scale of neutron scattering ($\sim$ 10$^{-10}$ seconds), are easily observed.

Wakimoto and co-workers made the remarkable observation in LSCO that the 2D IC AF structure rotates by 45$\degree$ at \textit{x} = 0.05 \cite{diagonalstripe}, exactly where superconducting ground states are first observed as a function of \textit{x}.  The same rotation \cite{dunsigerlbco} of the spin stripe structure was later observed in LBCO at the same doping level, \textit{x}, indicating that such a rotation of the 2D IC AF structure is a general feature of these single-layer cuprate systems.  The rotated stripes are referred to as ``parallel" spin stripes, as the hole-bearing domain walls are oriented along (100)$_{Tetragonal}$ directions, along Cu-O bonds \cite{Parallelstripe}.

On increasing the hole doping, the IC wavevector associated with the spin stripe structure increases $\sim$ \textit{x}, following what is known as the Yamada relation \cite{yamadarelation}.  This has the physical interpretation that the quasi-1D hole-bearing domain walls get closer together to allow for an increased hole density.  The onset temperature of this 2D IC AF order also rises strongly to a maximum near \textit{x} = 0.125  \cite{kofuhidden,hucker2011stripe}.  However, what happens at dopings beyond \textit{x} = 0.125 is less clear. This is the case for at least two reasons: First, the large single crystals necessary for definitive neutron scattering experiments are progressively more difficult to grow with increasing \textit{x}.  Second, the 2D magnetic signal becomes progressively weaker and spread out in \textbf{\textit{Q}}, due to smaller ordered moments and shorter correlation lengths associated with the the increased hole density.

This paper focuses on neutron scattering studies of the Nd-LSCO system. $\mu$SR and NMR techniques are sensitive to local magnetism on time scales much longer than those of neutron scattering, ($\sim$ 10$^{-8}$ seconds for $\mu$SR and $\sim$ 10$^{-6}$ seconds for NMR), and such studies have also been carried out on Nd-LSCO and other single-layer, hole-doped cuprates \cite{Nachumi,singer,hunt}.  While the energy scale for typical neutron diffraction ($\sim$ 1 meV) implies a time scale which is dynamic compared with both $\mu$SR and NMR, this elastic energy scale remains $\sim$ $1\over 2$ $\%$ or less of the full spin excitation bandwidth of these systems.  We henceforth refer to the corresponding magnetism elucidated by neutron scattering as static or quasi-static, but this proviso should be kept in mind, especially when comparing between different techniques.

Nd-LSCO has played a special role within this field, as Nd-LSCO \textit{x} = 0.125 was the first cuprate in which {\it both} parallel static spin stripe order and charge stripe order were observed \cite{tranquada1995evidence}.  Charge stripe order, with an incommensurate ordering wavevector twice that of the spin stripe order, is a natural consequence of the stripe picture, as it is the charge stripes that introduce a $\pi$ phase shift into the local $\pi$-$\pi$ antiferromagnetism.  At the \textit{x} = 0.125 doping level superconductivity is suppressed in Nd-LSCO to very low temperatures, with \textit{T}$_c$ $\sim$ 3 K \cite{tranquadaorderparameter,0p12tc}.  This phenomenon is known as the ``1/8 anomaly".  Nd-LSCO is also of interest as its maximum superconducting \textit{T}$_c$ is relatively low, $\sim$ 15 K \cite{michon2019thermodynamic}, compared with either LSCO, $\sim$ 40 K \cite{LSCOTc}, or LBCO, $\sim$ 35 K \cite{hucker2011stripe}.  This makes it easier to quench superconductivity with practical magnetic field strenghths in Nd-LSCO and to thereby explore the normal state properties at doping levels that would give rise to superconductivity at zero field.   

Such a deep suppression in superconducting \textit{T}$_c$ also occurs in LBCO, and co-existing static, parallel spin and charge stripes were later observed in this system around \textit{x} = 0.125, with static, parallel spin stripes observed to \textit{x} $\sim$ 0.135 \cite{hucker2011stripe}.  In contrast, LSCO displays a much milder suppression of \textit{T}$_c$ at \textit{x} = 0.125.  Both static parallel spin stripes and parallel charge stripes have been observed in LSCO, with static spin stripes observed to \textit{x} =  0.15 \cite{suzukisripelsco}, and parallel charge stripes observed in the vicinity of \textit{x} = 0.125 only \cite{croftcharge}. The combination of the relatively high onset temperature for charge stripes and an only-modestly-suppressed superconducting \textit{T}$_c$ at the $\frac{1}{8}$ anomaly in LSCO, has allowed the interaction between parallel charge stripes and superconductivity to be well studied.  The X-ray scattering intensity associated with the parallel charge stripes in LSCO around \textit{x} = 0.125 clearly diminishes on entering the superconducting state, providing compelling evidence for competition between superconductivity and charge stripes \cite{croftcharge}.

This body of work shows clearly that, while \textit{x} = 0.05 is a well defined minimum doping level for the appearance of static, parallel spin stripes in each of Nd-LSCO, LBCO, and LSCO, the extent of this parallel spin stripe phase and how it ends is much less clear.  Various phase diagrams for single-layer, hole-doped cuprates, show the static, parallel spin stripe phase to fade away by \textit{x} $\sim$ 0.135 for LBCO \cite{hucker2011stripe} and $\sim$ 0.15 for Nd-LSCO, with more limited data to \textit{x} $\sim$ 0.2 \cite{tranquadaorderparameter}.

In this paper, we present new elastic and inelastic neutron scattering measurements on four new single crystals of Nd-LSCO with \textit{x} = 0.125, 0.19, 0.24 and 0.26.  The \textit{x} = 0.26 doping level is close to the end of superconductivity in Nd-LSCO \cite{mirelakyle}, while the \textit{x} = 0.24 sample is close to the proposed quantum critical point, \textit{p}$^*$=0.23 $\pm$ 0.005, associated with the pseudo-gap phase in Nd-LSCO \cite{daou2009linear}.  These new measurements show well defined, but non-resolution-limited magnetic Bragg peaks to appear at low temperatures in all of these samples, significantly extending the range of concentrations over which static, parallel spin stripes have been observed in Nd-LSCO.  We report that the effective Cu ordered moment within the spin stripe structure at low temperatures drops off monotonically and steeply from \textit{x} = 0.125 to higher dopings, with a concomitant decrease in the in-plane correlation length.  This provides a reasonable explanation for why this signal has been so difficult to observe until now.  On the basis of these measurements we propose a new phase diagram for parallel spin stripes in Nd-LSCO which spans the Sr concentration range from 0.05 $\leq$ \textit{x} $\leq$ 0.26, completely overlapping with the regime over which a superconducting ground state is observed.  Consequently, the superconducting state at any \textit{x} in Nd-LSCO is entered from a state with extensive static, parallel spin stripe order.  We conclude that this parallel spin stripe state, in contrast with the parallel charge stripe state, does not compete with superconductivity in Nd-LSCO, it is a prerequisite for it. 

\begin{figure*}[hptb]
\hspace*{-0.3in}\linespread{1}
\includegraphics[width=7.2in]{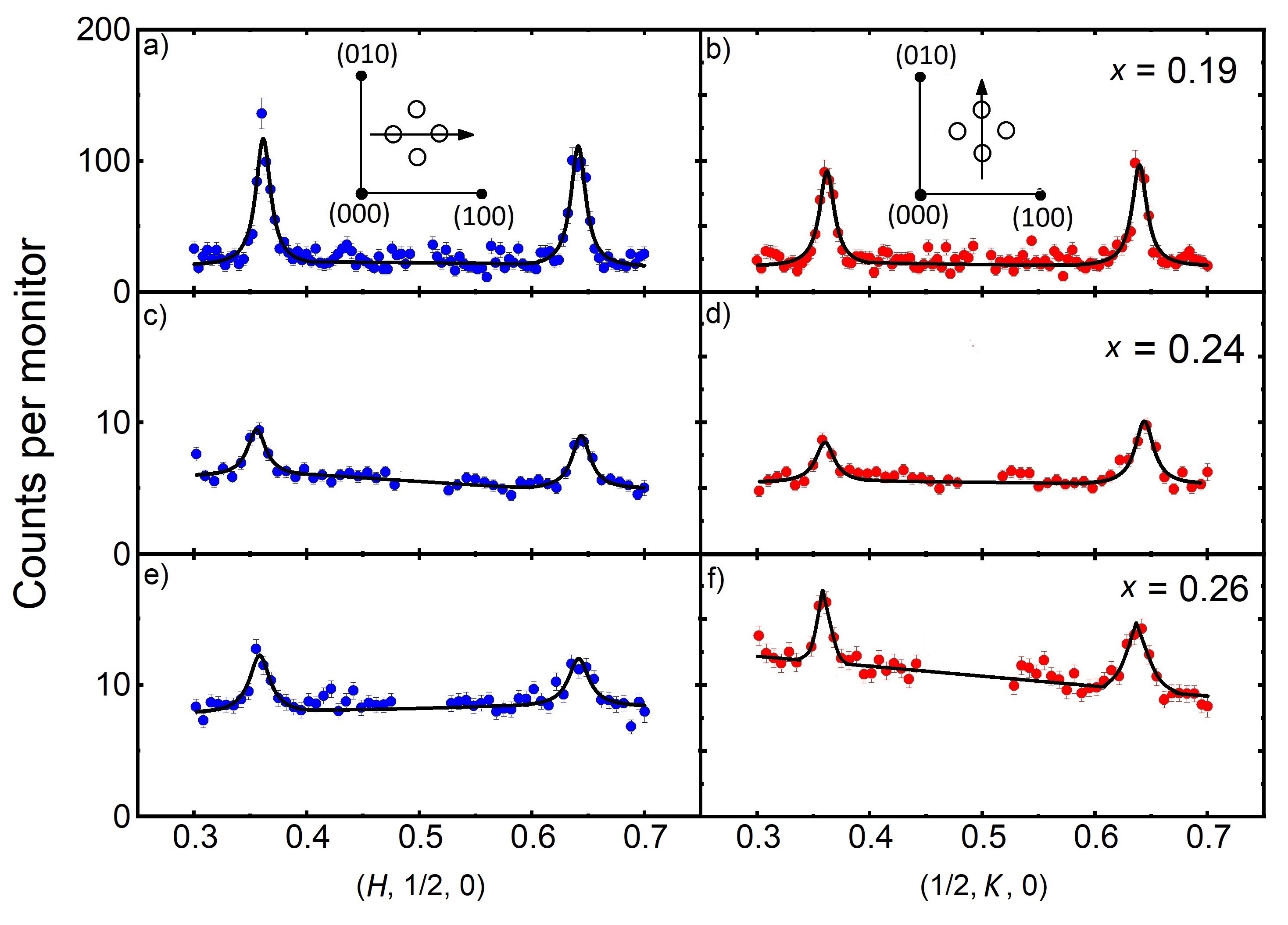}
\caption{Elastic TAS neutron scattering scans in reciprocal space for \textit{x} = 0.19, 0.24 and 0.26 single crystals of Nd-LSCO at \textit{T} = 1.5 K. The IC peaks can be observed in all three samples at reciprocal space positions ($\frac{1}{2}\pm\delta$, $\frac{1}{2}$, 0) and ($\frac{1}{2}$, $\frac{1}{2}\pm\delta$, 0) , $\delta$ $\approx$ 0.14. The insets to Fig. \ref{ICpeaks} a) and b) illustrate the \textit{H} and \textit{K} scans employed in reciprocal space.  The vertical axis displays neutron intensity,  counts per monitor for approximately 1\textit{s}. The black lines going through the data points are fits to the data described by Eq. 1.}
\label{ICpeaks} 
\end{figure*}

\section{Sample Preparation \& Experimental Methods}
High quality single crystals of Nd-LSCO with \textit{x} = 0.125, 0.19, 0.24 and 0.26 were grown using the traveling solvent floating zone technique at McMaster University with each resulting single crystal sample weighing between 3.5 and 5 grams. The single crystals were produced using a four-mirror Crystal Systems Inc. halogen lamp image furnace at approximate growth speeds of 0.68 mm/hr, and growths lasting for approximately 1 week each. The Sr concentration of each single crystal was determined by careful correlation of the structural phase transition temperatures with pre-existing phase diagrams, as described in \cite{mirelakyle}.  Further details regarding the materials preparation and single crystal growth of these samples, as well as determination of their stoichiometry is reported in \cite{mirelakyle}. The Nd-LSCO sample with \textit{x} = 0.19 was comprised of two co-aligned single crystals.  All single crystals were scanned using a back-scattering Laue instrument to assess their single-crystal nature.  Neutron diffraction measurements showed mosaic spreads of less than 0.5$\degree$ in all crystals, attesting to their high quality, single crystallinity.

Nd-LSCO single crystals with \textit{x} = 0.19, 0.26 were studied using the triple axis neutron spectrometer (TAS) Taipan at the Australian Centre for Neutron Scattering (ACNS), ANSTO, while the TAS experiment for the \textit{x} = 0.24 sample was conducted using the HB3 TAS instrument at the High Flux Isotope Reactor (HFIR) at Oak Ridge National Laboratory. Experiments using both Taipan and HB3 employed pyrolytic graphite monochromators, analysers and filters, and employed the same fixed final neutron energies: \textit{E}$_f$ = 14.7 meV.  Horizontal beam collimations of open-40$^\prime$-40$^\prime$-open for Taipan and 48$^\prime$-40$^\prime$-40$^\prime$-120$^\prime$ for HB3, were used.  The resulting energy resolution of the two sets of measurements was therefore similar, $\sim$ 0.9 meV.  For all TAS measurements the crystals were loaded in pumped $^4$He cryostats with a base temperature of 1.5 K.

Time-of-flight (TOF) neutron chopper spectrometer measurements were also carried out on the \textit{x} = 0.125, 0.19, and 0.24 single crystals using SEQUOIA at the Spallation Neutron Source, Oak Ridge National Laboratory \cite{granroth2010sequoia}.  These measurements were performed using \textit{E}$_i$=60 meV neutrons, which gave an energy resolution of $\sim$ 1.2 meV at the elastic position.  For all TOF measurements, the crystals were loaded in closed cycle refrigerators with a base temperature of 5 K.

\section{Experimental Results}

\subsection{Elastic neutron scattering}

Elastic neutron scattering scans of the form (\textit{H}, 1/2, 0) and (1/2, \textit{K}, 0) (using tetragonal notation) were carried out for the Nd-LSCO single crystals samples with \textit{x} = 0.19, 0.24 and \textit{x} = 0.26 with a base temperature \textit{T} = 1.5 K. These elastic scattering data are shown in Fig. \ref{ICpeaks}.  Four IC AF ($\frac{1}{2}\pm\delta$, $\frac{1}{2}$, 0) and ($\frac{1}{2}$, $\frac{1}{2}\pm\delta$, 0) quasi-Bragg peaks are observed, as expected for twinned orthorhombic structures.  Schematic trajectories of these scans in reciprocal space are shown in the insets to Fig. \ref{ICpeaks} a) and b). All the IC AF scattering is non-resolution-limited in \textbf{\textit{Q}} space, which is why we refer to these sharp diffraction features as quasi-Bragg peaks.  

\begin{figure}[tbp]
\linespread{1}
\hspace*{-0.2in} \includegraphics[width=3.5in]{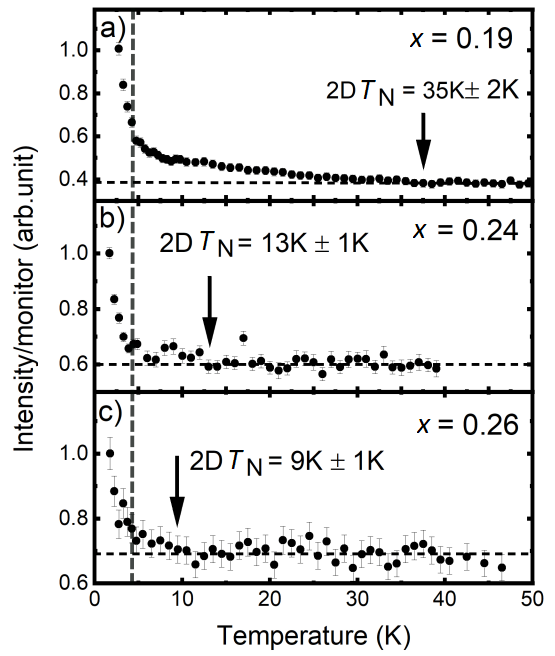}
\caption{Elastic, temperature scans on one of the 4 IC AF peaks at (0.5, 0.64, 0). The data were taken on warming the samples from base temperature $\sim$ 1.5 K up to 50 K, 40 K and 46 K with careful equilibration for each of the \textit{x} = 0.19, 0.24 and 0.26 samples, in a), b), and c), respectively.  The vertical dashed line indicates T=4 K, below which a pronounced upturn in all intensities is observed, due to Nd$^{3+}$-Cu$^{2+}$ coupling.  Horizontal dashed lines mark the high temperature background and the arrows indicate where the 2D IC AF Bragg scattering signal departs from the background in each sample. The estimated 2D \textit{T}$_N$ for \textit{x} = 0.19 is 35 K $\pm$2 K in a), for \textit{x} = 0.24 it is 13 K $\pm$ 1 K in b) and for \textit{x} = 0.26 it is 9 K $\pm$ 1 K in c).}
\label{Orderparameter}
\end{figure}

\begin{figure}[tbp]
\linespread{1}
\par
\hspace*{-0.2in}\includegraphics[width=3.6in]{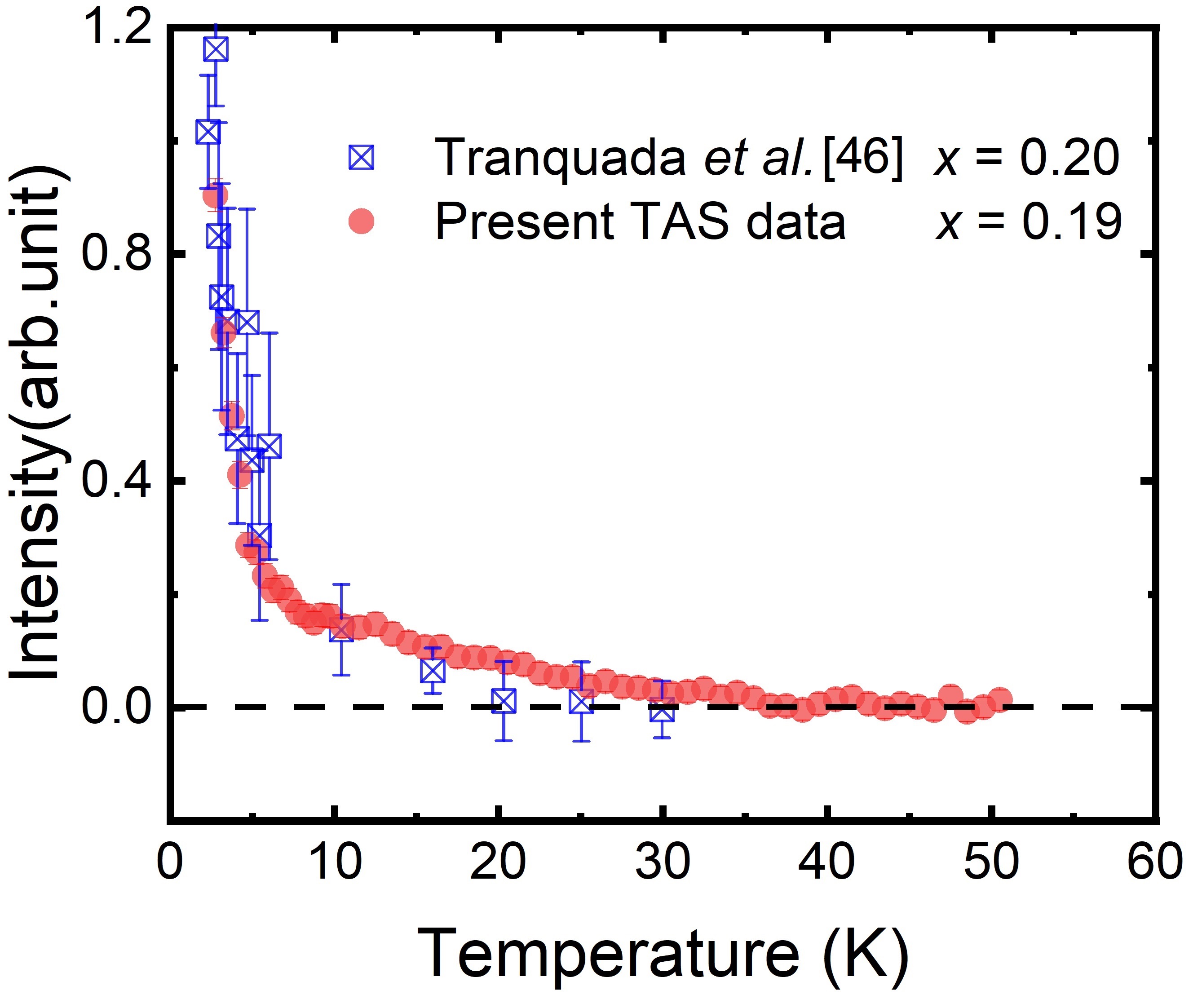}
\par
\caption{A comparison of the present TAS data order parameter on the sample \textit{x} = 0.19 sample and that reported for a \textit{x} = 0.20 sample \cite{tranquada1999glassy} are shown. The two elastic scattering data sets show very good agreement with each other. However the present data on the \textit{x} = 0.19 sample has much smaller error bars (smaller than data point) and much higher temperature point density, aiding in a more accurate estimate of 2D T$_N$ = 35 K $\pm$ 2 K.}
\label{kyleandtranquada} 
\end{figure}

These results show that static IC AF order exists at \textit{T} = 1.5 K in all three single crystals. The intensity for the \textit{x} = 0.19 sample is considerably stronger than that of \textit{x} = 0.24 or \textit{x} = 0.26, but the signals are otherwise qualitatively similar to each other.  They are also qualitatively similar to earlier elastic neutron scattering results on an \textit{x} = 0.125 sample by Tranquada \textit{et al.} \cite{tranquada1995evidence,tranquadaorderparameter,tranquada1999glassy,moment}. We therefore conclude that all Nd-LSCO single crystal samples with 0.125 $\leq$ x $\leq$ 0.26 display quasi-Bragg peaks at \textit{T} = 1.5 K which corresponds to IC AF order that is well defined in \textbf{\textit{Q}} and static on the time scale of the neutron scattering measurements ($\sim$ 10$^{-10}$ seconds).

The order parameters, or the temperature dependence of the Bragg intensities, were measured at the (0.5, 0.64,0) IC AF position, with careful thermal equilibration of the samples.  These order parameter measurements were performed for the Nd-LSCO single crystals with \textit{x} = 0.19, 0.24 and 0.26, as shown in Fig. \ref{Orderparameter} a), b), and c), respectively.  All three samples display a pronounced upturn in their intensity below $\sim$ 4 K, which is marked with a vertical fiducial in Fig. \ref{Orderparameter}.  Similar behaviour was also observed in the corresponding order parameter measurement of Nd-LSCO with \textit{x} = 0.125 by Tranquada \textit{et al.} \cite{tranquadaorderparameter}.  There, this upturn was successfully modelled as arising from the coupling between Cu$^{2+}$ and Nd$^{3+}$ moments and the onset of Nd$^{3+}$ magnetic correlations at low temperatures.  

\begin{figure}[tbp]
\linespread{1}
\par
\vspace*{-0.2in}\includegraphics[width=3.6in, height=6in]{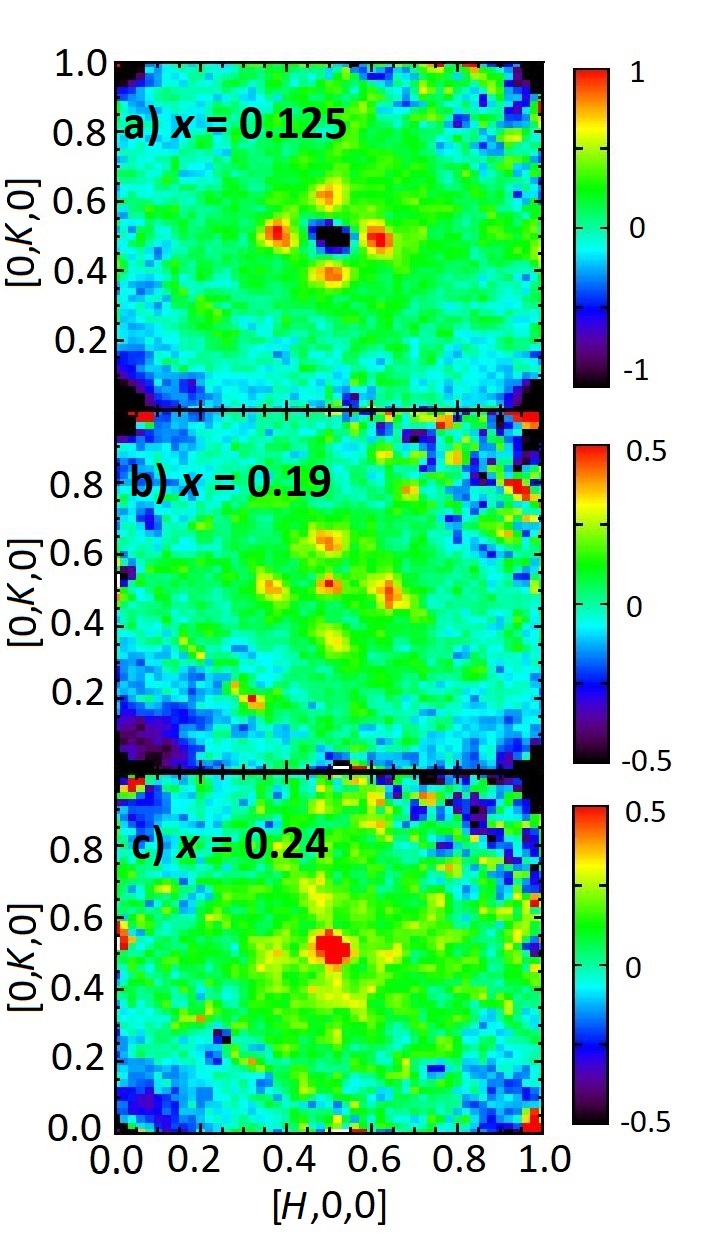}
\par
\caption{Elastic Time-of-flight (TOF) neutron data from Nd-LSCO with \textit{x} = 0.125, 0.19 and 0.24 in the [\textit{H}, \textit{K}, -4 $\leq$ \textit{L} $\leq$ 4] plane; that is with an integration in \textit{L} from -4 to 4 \textit{r.l.u}, in a), b), and c) respectively. These results show a subtraction of a high temperature data set from a low temperature data set, 5 K - 60 K, 5 K - 35 K, and 5 K - 35 K, respectively. The IC AF magnetic peaks around [1/2, 1/2, 0], arising from 2D parallel stripe order, are sharpest and most intense for \textit{x} = 0.125 in a), and gradually broaden and weaken with higher doping, 0.19 in b) and 0.24 in c). }
\label{sequoiafig} 
\end{figure}

This Cu$^{2+}$ - Nd$^{3+}$ coupling leads to the development of pronounced 3D correlations which peak at ordering wavevectors ($\frac{1}{2}\pm\delta$, $\frac{1}{2}$, 0) and ($\frac{1}{2}$, $\frac{1}{2}\pm\delta$, 0), that is with \textit{L} = 0.  At temperatures above $\sim$ 4 K, the IC AF Bragg scattering takes on a progressively more 2D nature, with scattering extended along \textit{L} in reciprocal space.  In addition, the coupling induces a rotation of the \textit{S} = 1/2 Cu$^{2+}$ moments out of the basal plane, so as to be parallel with the Nd$^{3+}$ moments, which themselves are largely constrained to be along \textit{c} due to crystal field effects.  As \textit{c} is perpendicular to the ($\frac{1}{2}\pm\delta$, $\frac{1}{2}$, 0) and ($\frac{1}{2}$, $\frac{1}{2}\pm\delta$, 0) ordering wavevectors, the intensity of the magnetic Bragg scattering also increases as a consequence of the Cu$^{2+}$ moments being more $\perp$ to \textbf{\textit{Q}}, which results in higher magnetic Bragg intensities due to the polarization dependence of the magnetic neutron scattering cross section. 

As the concentration of Nd is the same across this Nd-LSCO series, the same phenomenology is expected at all \textit{x}.  As can be seen from Fig. \ref{Orderparameter}, this is indeed the case with a pronounced factor of $\sim$ 3 to 4 enhancement in the strength of the IC AF Bragg scattering at ($\frac{1}{2}\pm\delta$, $\frac{1}{2}$, 0) and ($\frac{1}{2}$, $\frac{1}{2}\pm\delta$, 0) from \textit{T} $\sim$ 4 K to $\sim$ 1.5 K for all \textit{x}.  Of course, such an enhancement would not occur in the absence of Nd, hence the identification of the IC AF Bragg scattering at low temperatures is considerably easier in Nd-LSCO, as compared with LBCO or LSCO, especially at high doping.

The order parameter measurements also allow us to estimate the onset of the 2D IC antiferromagnetism, 2D \textit{T}$_N$.  From earlier work by Tranquada \textit{et al.} \cite{tranquadaorderparameter} on Nd-LSCO at \textit{x} = 0.125 and 0.15, 2D \textit{T}$_N$ is known to peak at $\sim$  50 K for \textit{x} = 0.125, where IC {\it charge} order is strongest, and the superconducting \textit{T}$_c$ is a local mimimum, $\sim$ 3 K.  Our order parameter data in Fig. \ref{Orderparameter}, shows 2D \textit{T}$_N$ to correspond to where the 2D IC AF Bragg scattering departs from a high temperature background.  The background employed is between 40 K $<$ T $<$ 50 K for \textit{x} = 0.19, 23 K $<$ \textit{T} $<$ 40 K for \textit{x} = 0.24, and 20 K $<$ \textit{T} $<$ 46 K for \textit{x} = 0.26. This gives estimates for 2D \textit{T}$_N$ of 35 K $\pm$ 2 K for \textit{x} = 0.19; 13 K $\pm$ 1 K for \textit{x} = 0.24; and 9 K $\pm$ 1 K for \textit{x} = 0.26, as marked by the arrows in Fig. \ref{Orderparameter} a), b) and c), respectively.

The same earlier work by Tranquada \textit{et al.} \cite{tranquadaorderparameter} on Nd-LSCO also measured the order parameter for an \textit{x} = 0.2 sample, very close in composition to our \textit{x} = 0.19 single crystal.  Figure 3 shows these two data sets overlaid on each other, and the raw data are clearly consistent with each other.  However, the error bars associated with the neutron intensity and the temperature point density associated with the new \textit{x} = 0.19 data is much improved over the pre-existing \textit{x} = 0.2 data, allowing us a better estimate of 2D \textit{T}$_N$ = 35 K $\pm$ 2 K, which we use in all subsequent discussion. 

\begin{figure}[htb]
\hspace*{-0.0in}\includegraphics[width=3.5in]{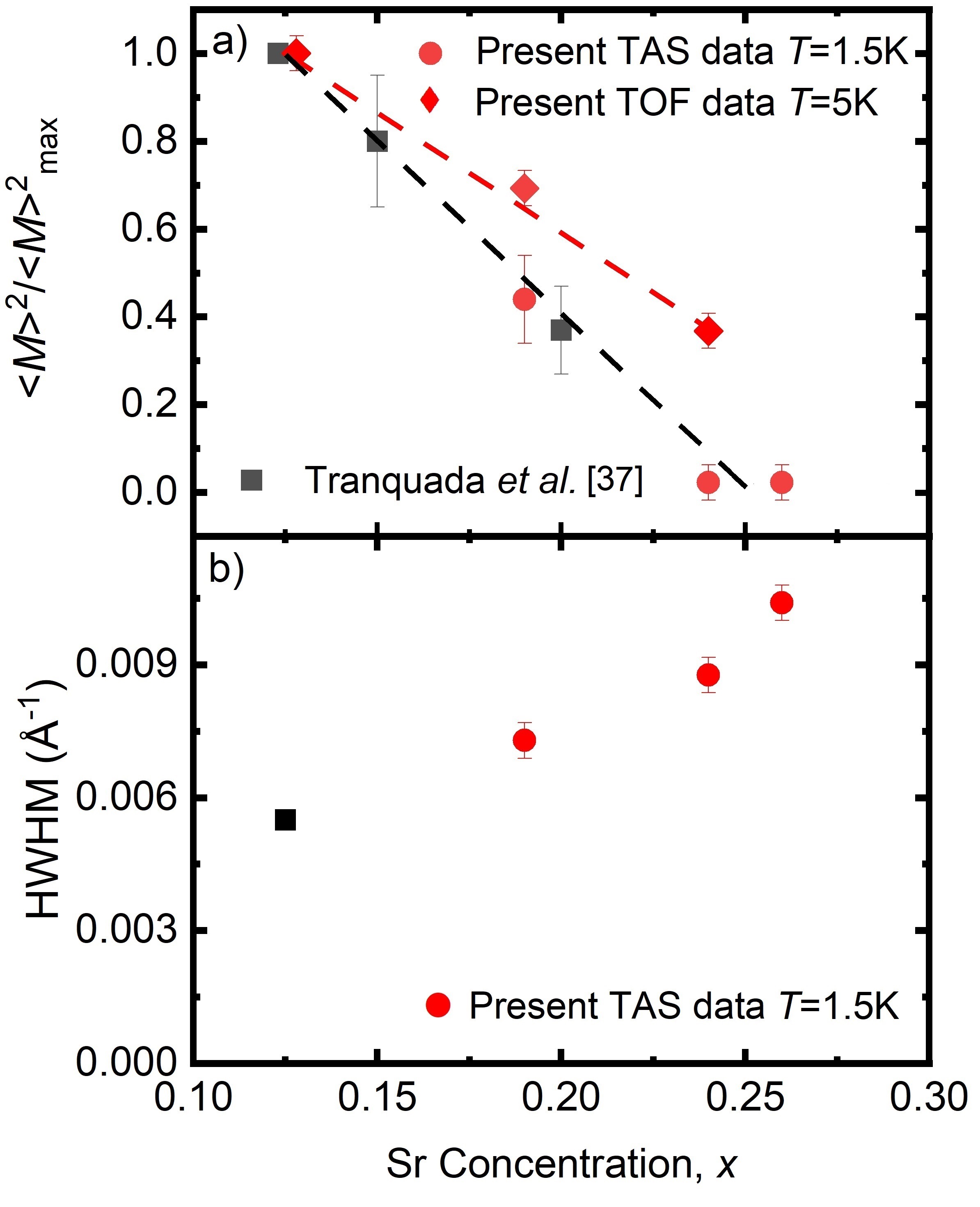}
\caption{a) The square of the sublattice magnetization at 1.5 K and 5 K, normalized to \textit{x} = 0.12. The black circles represent pre-existing data from Tranquada \textit{et al.} \cite{tranquadaorderparameter}. The red circles represent the current TAS data which shows consistency with the pre-existing data. The red diamonds represent the 5 K data taken as the integrated intensities of the IC AF peaks in Fig. \ref{sequoiafig} normalized to the $\sim$ 24 meV CEF levels in the system. b) The half-width at half-maximum of the current IC AF peaks in \textit{x} = 0.19, 0.24 and 0.26 are shown, along with pre-existing \textit{x} = 0.12 (black square) data from Tranquada \textit{et al.} \cite{tranquada1999glassy}}. 
\label{Mplot}
\end{figure}

TOF elastic neutron scattering data, complementary to the TAS data shown in Fig. \ref{ICpeaks}, are displayed within the (\textit{H}, \textit{K}, -4 $<$ \textit{L} $<$ 4) plane of reciprocal space, for \textit{x} = 0.125, 0.19 and 0.24 single crystals in Fig. \ref{sequoiafig}.  The \textit{x} = 0.19 and 0.24 single crystal samples are the same single crystal samples used for the TAS measurements in Fig. \ref{ICpeaks}.  These three concentrations correspond to Nd-LSCO with peak onset \textit{T}$_N$ for 2D IC AF order, optimal doping for superconductivity, and close to the end of the superconducting dome, respectively.  All three data sets show a difference between \textit{T} = 5 K data and that taken at a temperature equal to or above 2D \textit{T}$_N$ (\textit{T} = 60 K for \textit{x} = 0.125 and \textit{T} =35 K for each of \textit{x} = 0.19 and 0.24). This data was extracted from four dimensional (three \textbf{\textit{Q}} dimensions, as well as energy) data sets taken with the TOF chopper spectrometer, SEQUOIA. The elastic scattering data are integrated in energy over the range -2 meV $<$ \textit{E} $<$ 2 meV, which encompasses the $\sim$ 1.2 meV energy resolution for SEQUOIA measurements employing \textit{E}$_i$ = 60 meV.  The integration over -4 $<$ \textit{L} $<$ 4 is necessary to capture sufficient 2D IC AF Bragg intensity, which is extended along \textit{L} at \textit{T} = 5 K, and concentrated at relatively low \textbf{\textit{Q}}, due to the magnetic form factor of Cu$^{2+}$.  Allowed nuclear Bragg peaks occur in these orthorhombic structures at wavevectors such as ($\frac{1}{2}$, $\frac{1}{2}$, \textit{L} = 0, $\pm$1, $\pm$2, $\pm$3), using tetragonal notation.  Hence relatively strong nuclear intensity is observed at these positions. These structural zone centres also have low energy acoustic phonons associated with them, whose intensity increases with increasing temperature, and which, while weak, can be picked up within the energy integration performed here.  As a result, the temperature difference plots shown in Fig. \ref{sequoiafig} have relatively strong positive or negative intensity at the ($\frac{1}{2}$, $\frac{1}{2}$, -4 $<$ \textit{L} $<$ 4) position, due to the subtraction of two relatively large nuclear intensities.  This is an artifact of the subtraction between data sets at two different temperatures. 

These measurements were performed on three single crystals of similar size and under very similar conditions.  Although a normalization of the strength of the IC AF Bragg intensities is straightforward, and will be carried out below, it is useful to consider the data directly.  We can see that the IC AF Bragg intensity at \textit{T} = 5 K is well defined and relatively strong for \textit{x} = 0.125, becoming substantially weaker for \textit{x} = 0.19, and then both weak and broad for \textit{x} = 0.24.

An analysis of the elastic TAS data in Fig. \ref{ICpeaks}, and the elastic TOF data in Fig. \ref{sequoiafig}, was carried out so as to quantitatively estimate the relative strength of the IC AF Bragg scattering as a function of \textit{x} for our \textit{x} = 0.125, 0.19, 0.24 and 0.26 samples at both \textit{T} = 1.5 and 5 K.  The \textit{H} and \textit{K}, elastic TAS scans in Fig. \ref{ICpeaks} were fit to Lorentzian lineshapes of the form 

\begin{equation}
S(\textbf{\textit{Q}}, \hbar\omega=0)=\dfrac{I_0}{(\textbf{\textit{Q}}-\textbf{\textit{Q}}_0)^2 + (\dfrac{1}{2}\Gamma)^2},
\end{equation}

(where $\Gamma$ indicates the full width at half maximum of the function and $\textbf{\textit{Q}}_0$ is its center) with the purpose of extracting both the integrated intensity of the IC AF Bragg scattering and its in-plane correlation length.  The intensity of the IC AF Bragg scattering was normalized against rocking curve measurements of the intensity of structurally allowed (1,1,0) Bragg peaks, such that the intensities of the IC AF Bragg scattering at \textit{T} = 1.5 K could be quantitatively compared between the \textit{x} = 0.19, 0.24 and 0.26 samples.

A similar normalization protocol was followed for the IC AF Bragg scattering in the \textit{x} = 0.125, 0.19 and 0.24 samples at \textit{T} = 5 K, using the TOF data in Fig. \ref{sequoiafig}.  As this elastic scattering data is part of much larger inelastic scattering data sets, we can normalize the IC AF integrated Bragg scattering to that of crystalline electric field (CEF) excitations near $\sim$ 24 meV.  A study on these CEF excitations will be reported separately, but their intensity depends primarily on the amount of Nd present in the samples, which is the same in all cases.  

These results are summarized in Fig. \ref{Mplot} a) for the integrated intensities and Fig. \ref{Mplot} b) for the correlation lengths.  In both cases our new results are put into the context of previous measurements on Nd-LSCO single crystal samples \cite{tranquadaorderparameter,tranquada1999glassy}. The primary conclusions are that the \textit{x}-dependence of the intensity of the IC AF Bragg peaks, which goes like the square of the sublattice magnetization, $<\textit{M}>^2$, is consistent between our data at \textit{T} = 1.5 K on \textit{x} = 0,19, 0.24, and 0.26 samples, and that of Tranquada \textit{et al.} on \textit{x} = 0.125, 0.15, and 0.20 samples  \cite{tranquadaorderparameter}.  At this low temperature, \textit{T} = 1.5 K, the influence of the Cu$^{2+}$-Nd$^{3+}$ coupling is profound and the scattering has significant 3D correlations.  In contrast our TOF data for \textit{x} = 0.125, 0.19, and 0.24 samples at \textit{T} = 5 K, is much more 2D in nature.  We can see that this TOF integrated static magnetic intensity also falls off sharply with \textit{x} from \textit{x} = 0.125 to 0.24, but at about half the rate seen for the TAS results at \textit{T} = 1.5 K.  We attribute this difference between the \textit{x}-dependence of the static magnetism at \textit{T} = 1.5 K and that at \textit{T} = 5 K, to an \textit{x}-dependence in the nature of the effectiveness of the coupling between Cu$^{2+}$ and Nd$^{3+}$ magnetic moments.  We conclude that the coupling is stronger and thus more efficient for smaller \textit{x}, where more Cu$^{2+}$ moment is present.  

Earlier work by Tranquada \textit{et al.} on \textit{x} = 0.125 concluded that the ordered Cu$^{2+}$ moment at low temperatures was 0.1 $\mu_B$ $\pm$0.03 $\mu_B$ \cite{moment}. Our new work then suggests that the ordered Cu$^{2+}$ moment participating in the 2D IC AF structure at \textit{T} = 5 K in our \textit{x} = 0.24 single crystal samples is roughly 40$\%$ smaller, thus $\sim$ 0.06 $\mu_B$.

The in-plane magnetic correlation lengths at \textit{T} = 1.5 K are plotted in Fig. \ref{Mplot} b) for our new single crystals \textit{x} = 0.19, 0.24, and 0.26.  Previous studies on IC AF order for \textit{x}=0.125 had already determined that the relevant magnetic Bragg features had finite in plane correlation length, in spite of the fact that the IC AF order is at its strongest at this doping level \cite{moment}.  Our new results are consistent with those for \textit{x} = 0.125, in that the sequence of low temperature in-plane correlation lengths is seen to decrease monotonically with doping (the half-width at half-maximum increases monotonically), and the overall trend from \textit{x} = 0.125 to 0.26 is an $\sim$ linear relationship.


\subsection{Low Energy Inelastic Magnetic Scattering; \\ Inelastic and Elastic Order Parameters}
\label{subsec:cf-analysis}

\begin{figure}[tbp]
\linespread{1}
\par
\vspace*{-0.16in}\hspace*{-0.2in}\includegraphics[width=3.5in]{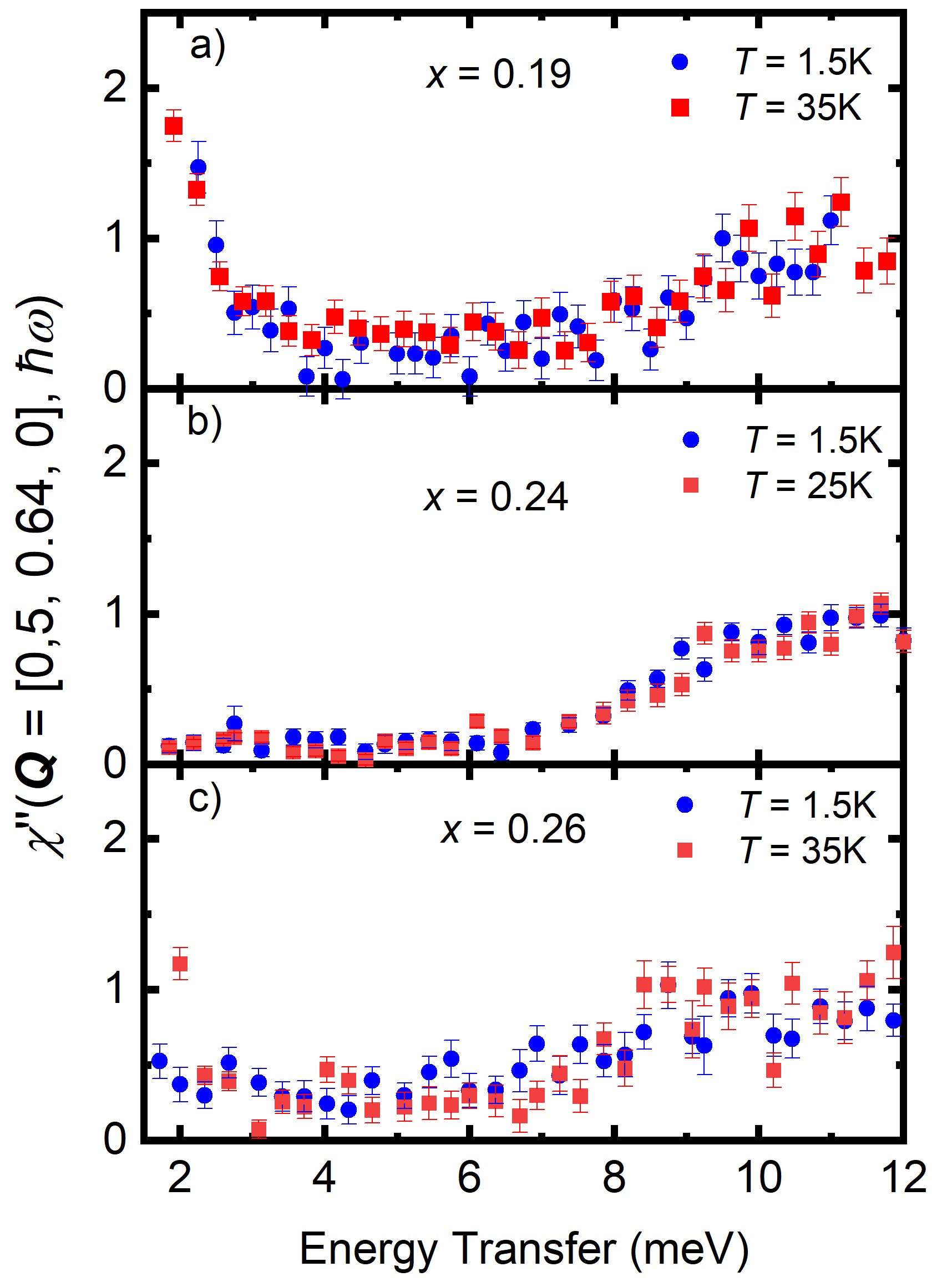}
\par
\caption{Energy scans of $\chi \prime\prime (\textbf{\textit{Q}}, \hbar\omega)$ taken from TAS constant-Q measurements at \textbf{\textit{Q}}=[0.5, 0.64, 0] - one of the four incommensurate AF peak positions around the [0.5, 0.5, 0] position.}
\label{energyscan}
\end{figure}

\begin{figure}[tbp]
\linespread{1}
\vspace*{-0.2in}\includegraphics[width=3.4in]{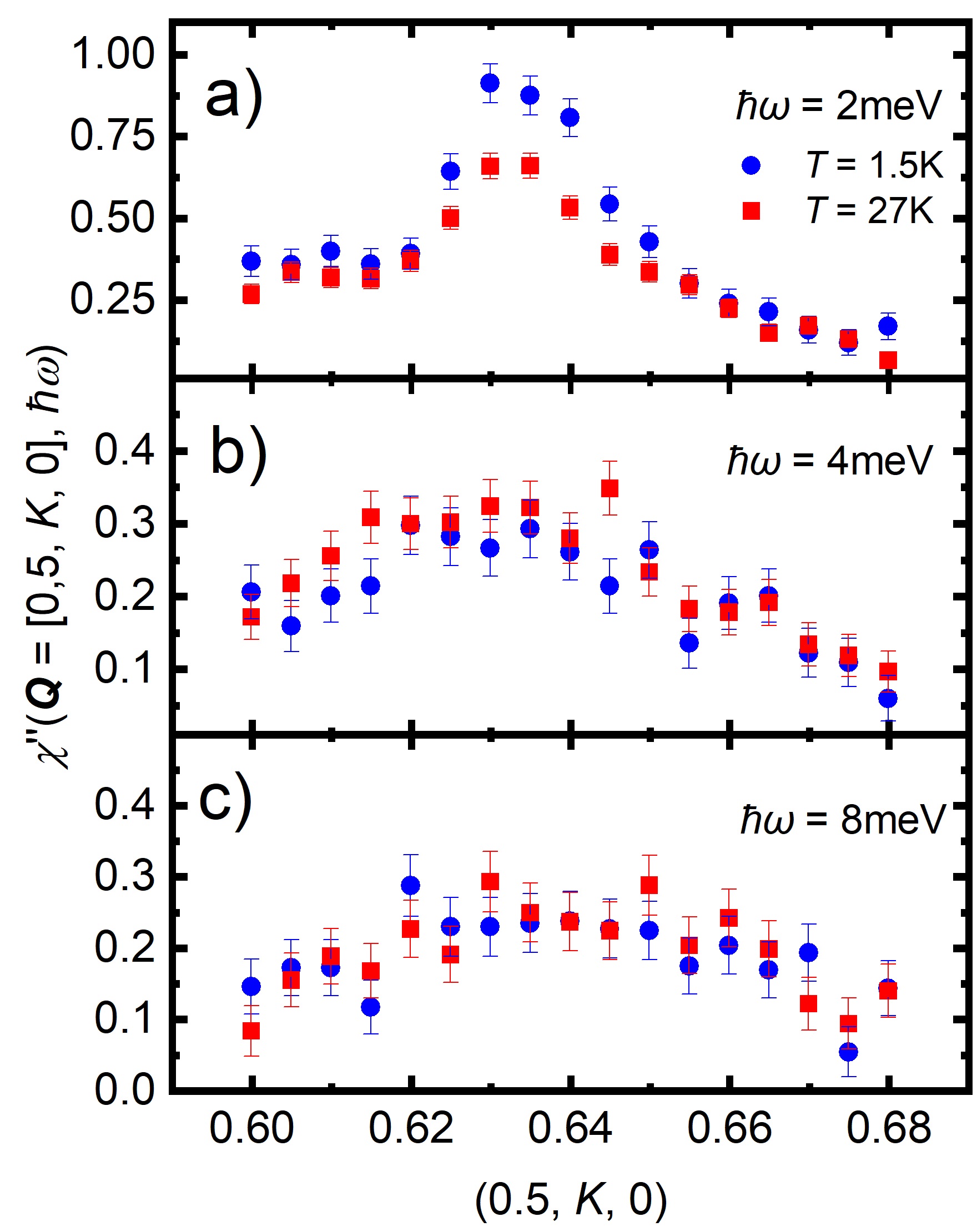}
\caption{$\chi \prime\prime (\textbf{\textit{Q}}, \hbar\omega)$ is shown for \textbf{\textit{Q}} = [0.5, 0.64, 0] and $\hbar\omega$ = 2 meV, 4 meV and 8 meV in a), b) and c), respectively, from TAS constant energy transfer ($\hbar\omega$) scans at both low (\textit{T} = 1.5 K) and high temperatures (\textit{T} = 27.5 K).}
\label{ecuts}
\end{figure}

The low energy spin fluctuations within the static, parallel spin stripe state of Nd-LSCO were also investigated using TAS spectroscopy.  Constant-\textbf{\textit{Q}} measurements were performed at the IC AF zone centre, (0.5, 0.64, 0), as a function of energy transfer, as shown for \textit{x} = 0.19, 0.24 and 0.26 in Fig. \ref{energyscan} a), b) and c) respectively.  Constant energy scans were also performed as a function of \textbf{\textit{Q}}, across an IC AF zone centre with scans of the form (0.5, \textit{K}, 0), as shown in Fig. \ref{ecuts} for the \textit{x} = 0.19 single crystal.

\begin{figure}[tp]
\linespread{1}
\hspace*{-0.2in}\includegraphics[width=3.5in]{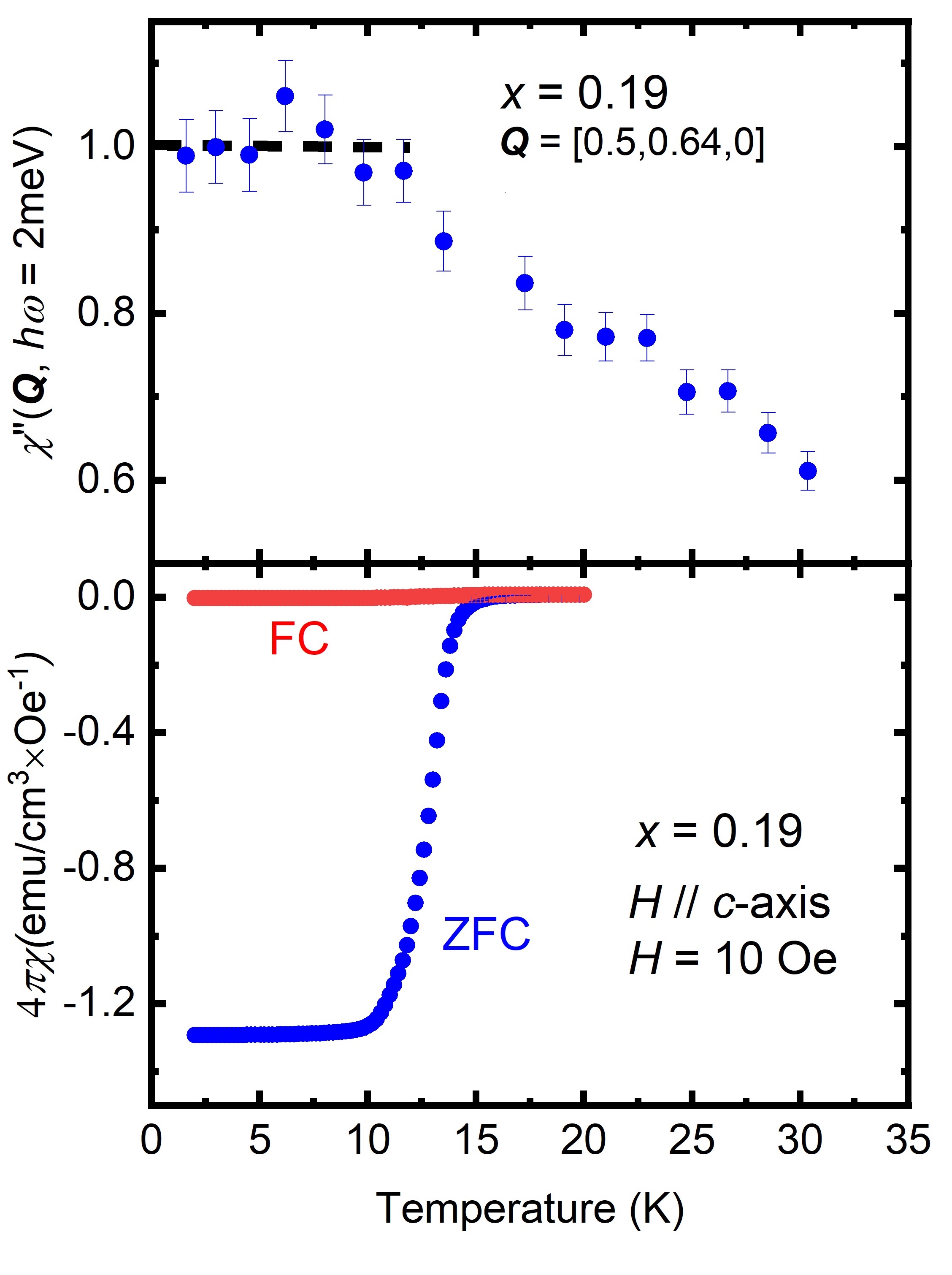}
\caption{a) The ``inelastic order parameter", $\chi \prime\prime (\textbf{\textit{Q}}, \hbar\omega)$ as a function of temperature for \textbf{\textit{Q}} = [0.5, 0.64, 0] and $\hbar\omega$ = 2 meV for Nd-LSCO with \textit{x} = 0.19.  These data were derived from TAS measurements. The horizontal dashed line indicates the plateau behaviour for \textit{T} $\leq$ \textit{T}$_c$ $\sim$ 13.5 K. b) Field cooled and zero field cooled magnetic susceptibility measurements on a small single crystal of the \textit{x} = 0.19 Nd-LSCO sample with a magnetic field of 10 \textit{Oe} applied parallel to \textit{c}-axis show T$_C$ $\sim$ 13.5 K.}
\label{insorderp}
\end{figure}

Figures \ref{energyscan}, \ref{ecuts}, and \ref{insorderp} a) show the imaginary part of the dynamic susceptibility, $\chi \prime\prime (\textbf{\textit{Q}}, \hbar\omega)$, as a function of both $\hbar\omega$ and \textbf{\textit{Q}}, respectively. This quantity is related to the inelastic neutron scattering intensity, the dynamic structure factor, S$(\textbf{\textit{Q}}, \hbar\omega)$, by: 

\begin{equation}
S(\textbf{\textit{Q}}, \hbar\omega)=[1-\exp({-\hbar\omega\over {k_BT}})]^{-1}\times \chi \prime\prime (\textbf{\textit{Q}}, \hbar\omega), 
\end{equation}

Isolating $\chi \prime\prime (\textbf{\textit{Q}}, \hbar\omega)$ from the dynamic structure factor, $S(\textbf{\textit{Q}}, \hbar\omega)$, involves making a robust estimate of the background, and then dividing the inelastic neutron scattering signal by the Bose factor, $[1-exp({-\hbar\omega\over {k_BT}})]^{-1}$.  This has been done for the data shown in Figs. \ref{energyscan}, \ref{ecuts} and \ref{insorderp}, using the scattering away from IC AF Bragg positions, such as (0.5, 0.68, 0), to estimate the background level.  


Inspection of the low energy dependence of $\chi \prime\prime (\textbf{\textit{Q}}, \hbar\omega)$ at \textbf{\textit{Q}} = (0.5, 0.64, 0) in Fig. \ref{energyscan} shows that this spectral weight is qualitatively different for the optimally-doped \textit{x} = 0.19 sample, with superconducting \textit{T}$_c$ $\sim$ 14 K, as compared with either of the \textit{x} = 0.24 or 0.26 single crystals which lie close to the end of the superconducting dome.  While all three samples show pronounced inelastic scattering which peaks up near a low lying CEF excitation at $\sim$ $\hbar\omega$ = 11 meV, only the \textit{x} = 0.19 sample shows strong quasi-elastic scattering which rises up in intensity below $\sim$ 4 meV.  For the higher-doped samples, \textit{x} = 0.24 and 0.26, very little inelastic magnetic spectral weight is obvious at energies less than $\sim$ 4 meV in these measurements.

More detailed constant-energy \textbf{\textit{Q}}-scans show that this low energy spectral weight is strongly peaked at the IC AF ordering wavevector, as can be seen in Fig. \ref{ecuts} for \textit{x} = 0.19. A strong peak is seen in the $\chi \prime\prime (\textbf{\textit{Q}}, \hbar\omega = 2 meV)$ data, that is not seen in the corresponding data at higher energies, $\hbar\omega$ = 4 or 8 meV.  At $\hbar\omega$ = 2 meV, the strength of $\chi \prime\prime (\textbf{\textit{Q}}, \hbar\omega)$ decreases by $\sim$ 40$\%$ on warming from \textit{T} =1.5 K to \textit{T} = 27 K. $\chi \prime\prime (\textbf{\textit{Q}}, \hbar\omega)$ at $\hbar\omega$ = 4 or 8 meV remains peaked around the IC AF ordering wavevector, but it is both considerably broader in \textbf{\textit{Q}} and weaker in intensity than that at 2 meV.  As expected, $\chi \prime\prime (\textbf{\textit{Q}}, \hbar\omega)$ at $\hbar\omega$ = 4 or 8 meV shows little or no temperature dependence to 27 K.

We have measured the temperature dependence of the IC AF inelastic spectral weight  at $\hbar\omega$ = 2 meV for the \textit{x} = 0.19 sample.  This is shown in Fig. \ref{insorderp} a), where $\chi \prime\prime (\textbf{\textit{Q}} = (0.5, 0.64, 0), \hbar\omega = 2 meV)$ is shown as a function of temperature from \textit{T} = 1.5 K to \textit{T} = 30 K.  This data can be thought of as an ``inelastic order parameter", and it is interesting to compare it to the actual order parameter, that is the elastic scattering measured at the same IC AF wavevector shown in Fig. \ref{ICpeaks} and \ref{Orderparameter}.  The order parameter for \textit{x} = 0.19, Fig. \ref{Orderparameter} a), shows that the 2D \textit{T}$_N$ = 35 $\pm$ 2 K.  Furthermore, it rises sharply with decreasing temperature at low temperatures, due to coupling between the Cu$^{2+}$ and Nd$^{3+}$ moments which induces substantial 3D spin correlations.  However the spin fluctuations captured in $\chi \prime\prime (\textbf{\textit{Q}}, \hbar\omega = 2 meV)$ at the IC AF wavevector rise linearly with decreasing temperature from 30 K to $\sim$ 13 K, before leveling off at low temperatures.  

This is a very interesting result as it shows the spectral weight of these low energy spin fluctuations in this optimally-hole-doped cuprate to saturate at temperatures below the approximate superconducting \textit{T}$_c$ of this single crystal.  In this \textit{x} = 0.19 sample, the temperatures of relevance for 2D magnetic order (2D \textit{T}$_N$ $\sim$ 35 K) and strong 3D correlations ($\sim$ 4 K) are well separated from that associated with superconductivity. For ease of reference, the measured magnetic susceptibility of a small piece of the single crystal used in neutron scattering experiments is plotted in Fig. \ref{insorderp} b) directly below the $\chi \prime\prime (\textbf{\textit{Q}}, \hbar\omega)$ temperature dependence data for $\hbar\omega$ = 2 meV.  A strong diamagnetic signal is observed with an onset near \textit{T}$_c$ $\sim$ 14.5 K, but the midpoint on the zero-field-cooled diamagnetic susceptibility curve occurs at $\sim$ 13.5 K, coincident with the start of the plateau in the inelastic, $\hbar\omega$ = 2 meV spectral weight at low temperatures.  A similar trend in the temperature dependence of $\chi \prime\prime (\textbf{\textit{Q}}, \hbar\omega)$ at low energies has been reported for Nd-LSCO with \textit{x} = 0.15 \cite{tranquada1999glassy}, but with a lower temperature-point-density, making it hard to identify characteristic temperatures.

\begin{figure}[btp]
\linespread{1}
\par
\vspace*{-0.1in}\hspace*{-0.2in}\includegraphics[width=3.4in,height=3in]{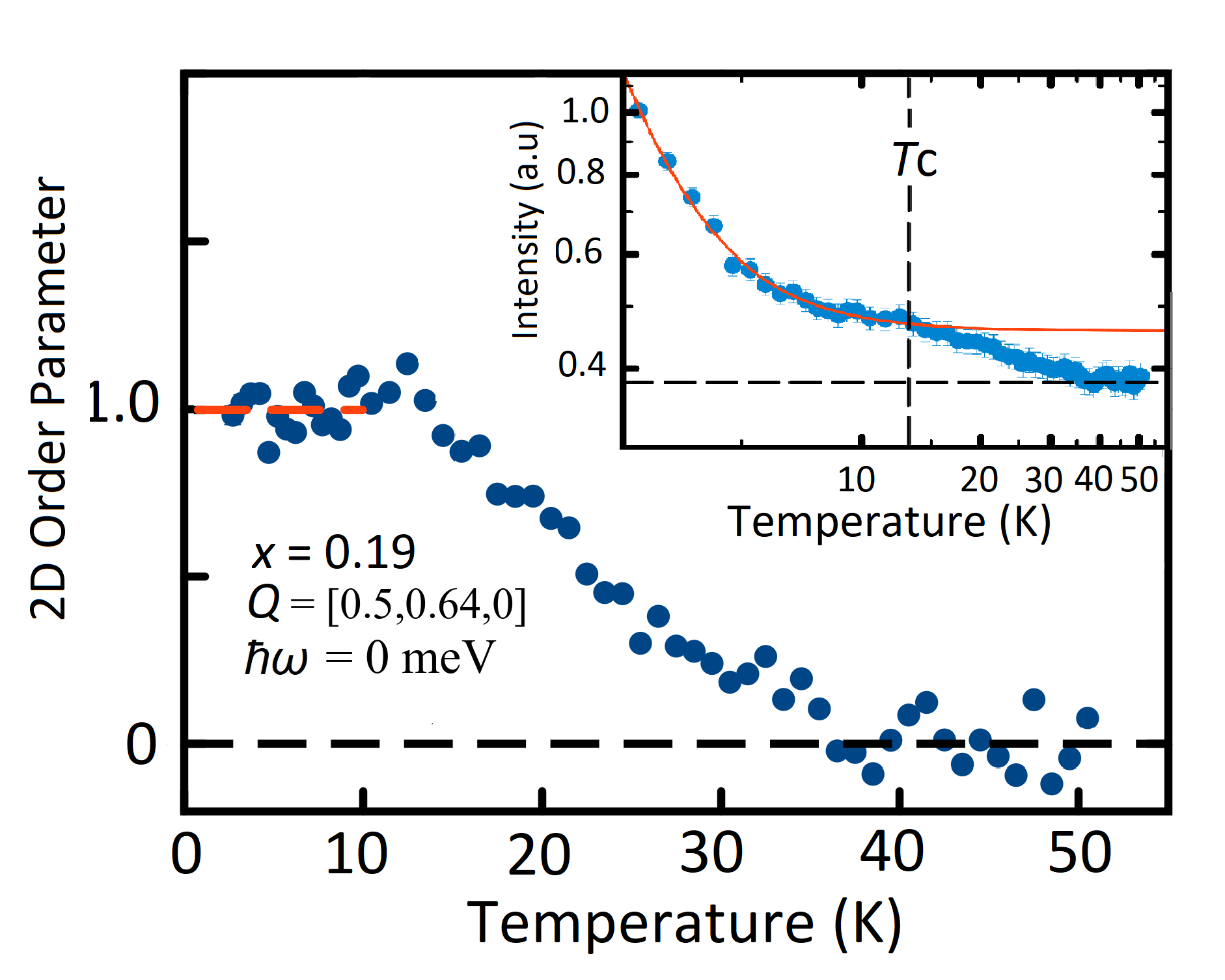}
\caption{The 2D Order Parameter (2DOP) for Nd-LSCO \textit{x} = 0.19 is shown. This is derived from the order parameter (OP) measurements shown in the inset and in Fig. \ref{Orderparameter} a), using the relation $OP(T)=A(T) \times 2DOP(T)$.  The amplification factor, A(\textit{T}), is derived from fitting the OP(T) at \textit{T} $\leq$ 12 K, as shown with the red line in the inset.  Like the ``inelastic order parameter" in Fig. \ref{insorderp}, the 2DOP displays the onset of a low temperature plateau coincident with \textit{T}$_c$. The order parameter shown in the inset is the same data as in Fig. \ref{Orderparameter} a), but on a log-log scale.  Plateau like behavior for \textit{T} $<$ \textit{T} $_C$ $\sim$ 13.5 K (indicated by the vertical dashed fiducial) is evident in this plot of the elastic neutron scattering order parameter itself.}
\label{2DOP}
\end{figure}

Close examination of the elastic order parameter for \textit{x} = 0.19, also shows evidence for a plateau below \textit{T}$_c$ $\sim$ 13.5 K.  The inset to Fig. \ref{2DOP} reproduces the order parameter from Fig. \ref{Orderparameter} a), but now on a log log scale.  The high temperature background (averaged between 40 K and 50 K) is indicated by the horizontal dashed lines in the inset.  We have modelled the strong upturn in intensity at low temperatures using a phenomenological expression 
\begin{equation}
A(T)=a + bT^{-\alpha}
\end{equation}

for temperatures $\leq$ 12 K, as shown by the dashed red line in the inset to Fig. \ref{2DOP}. We then decompose our measured order parameter ($OP$) according to 
\begin{equation}
OP(T)=A(T) \times 2DOP(T),    
\end{equation}

which gives the two dimensional order parameter ($2DOP$) for our \textit{x} = 0.19 sample in the main panel of Fig. \ref{2DOP}. This decomposition has the interpretation that the effect of the Cu$^{2+}$ - Nd$^{3+}$ coupling is to amplify the preexisting 2D Cu$^{2+}$ order parameter at low temperatures.  While it is phenomenological, it reproduces in the elastic order parameter, the same plateau behavior as seen in the temperature dependence of $\chi \prime\prime (\textbf{\textit{Q}}, \hbar\omega = 2 meV)$, shown in Fig. \ref{insorderp}.

In fact, A(\textit{T}) is rather insensitive to the precise range of temperatures $\leq$ 15  K that is used to fit the OP data.  As its amplification effect is rather minor until temperatures less than $\sim$ 8 K, evidence for the plateau can also be seen in the raw elastic scattering data in the inset to Fig. \ref{2DOP}, where superconducting \textit{T}$_C$ $\sim$ 13.5 K is indicated with a dashed fiducial reference. Additionally, as A(\textit{T}) is $\sim$ constant for T greater than $\sim$ 12 K, the form of the OP and the 2DOP are identical for temperatures above $\sim$ 1/3 $\times$ 2D \textit{T}$_N$.

Such intriguing plateaus as a function of decreasing temperature in the magnetic Bragg intensity and low energy spectral weight of the spin fluctuations has been previously observed in several heavy fermion superconductors, for example UPt$_3$  \cite{aeppli1988magnetic,Issac} and UPd$_2$Al$_3$ \cite{metoki1997coupling}.  In both cases, the coincidence of the plateaus with the superconducting \textit{T}$_c$ was taken as evidence for strong coupling between magnetism and superconductivity and a foundational role played by AF spin fluctuations in mediating superconductivity in these highly-correlated, heavy fermion metals.

\section{Phase Diagram and Discussion}

Our measurements of 2D \textit{T}$_N$, the onset of 2D parallel stripes at optimal and high hole-doping levels in Nd-LSCO, with \textit{x} = 0.19, 0.24, and 0.26, allow us to complete the phase diagram for 2D parallel stripes and to examine their relation to superconductivity.  These results extend earlier measurements for Nd-LSCO with \textit{x} $\leq$ 0.15 \cite{tranquadaorderparameter,wakimotondlsco}.  We have complemented these Nd-LSCO data with corresponding results from LBCO for 0.05 $\leq$ \textit{x} $\leq$ 0.125 \cite{dunsigerlbco,dunsigerlbco2,hucker2011stripe}.  LBCO is a useful proxy for Nd-LSCO at these low doping values, as both LBCO and Nd-LSCO show very pronounced ``$\frac{1}{8}$ anomalies" with superconducting \textit{T}$_c$'s suppressed to $\sim$ 3 K \cite{Axelbco}. In addition, the onset temperature for 2D parallel stripes in both LBCO and Nd-LSCO are maximal at \textit{x} = 0.125, with similar 2D \textit{T}$_N$ s $\sim$ 50 K \cite{tranquadaorderparameter,hucker2011stripe}.

\begin {figure}[tb]
\linespread{1}
\par
\hspace*{-0.2in}\vspace*{-0.2in}\includegraphics[width=3.5in]{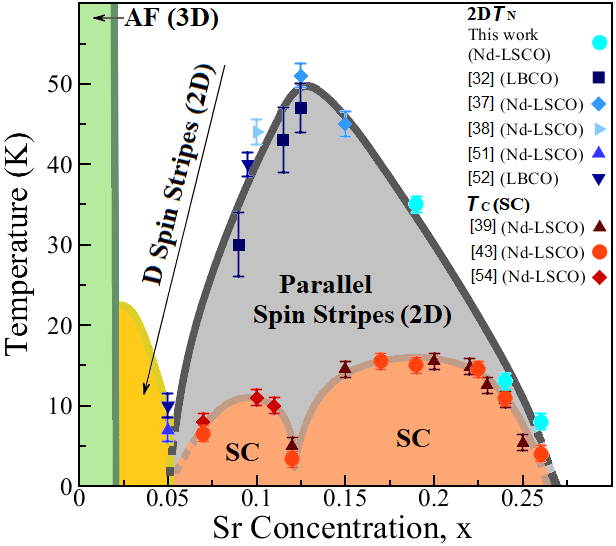}
\par
\caption{Phase diagram for quasi-static magnetism in Nd-LSCO as measured by neutron scattering. Combining the current TAS data for 2D \textit{T}$_N$ with pre-existing Nd-LSCO data, and complemented with LBCO results for 0.05 $\leq$ \textit{x} $\leq$ 0.125, we extend the regime of a 2D parallel spin stripe phase to 0.05 $\leq$ \textit{x} $\leq$ 0.26. Such a 2D parallel spin stripe phase is then a prerequisite for superconductivity. All 2D \textit{T}$_N$ parallel spin stripe data comes from elastic neutron scattering studies of either Nd-LSCO or LBCO as indicated in the legend. Superconducting \textit{T}$_C$ for Nd-LSCO are also from references indicated in the legend.}
\label{phasediagram}
\end{figure}

Fig. \ref{phasediagram} shows the phase diagram for 2D parallel stripes in Nd-LSCO, superposed with its phase diagram for superconductivity.  The superconducting transition temperatures are taken from earlier work \cite{axe1994structural,michon2019thermodynamic,mirelakyle}, and includes data for the four single crystal samples that were the subject of the present neutron scattering experiments, \textit{x} = 0.125, 0.19, 0.24 and 0.26 \cite{mirelakyle}.  As previously mentioned, the superconducting phase behavior shows an onset of superconducting ground states at \textit{x} = 0.05, which coincides with the rotation of the 2D diagonal spin stripe structure to the 2D parallel spin stripe structure.  Optimal superconducting \textit{T}$_c$ $\sim$ 15 K is reached for Nd-LSCO at x$\sim$ 0.19, and the end of the superconducting dome occurs near x$\sim$ 0.27 \cite{mirelakyle}.

We can see that the entire region of superconducting ground states for Nd-LSCO is contained within the 2D parallel stripe phase; superconductivity begins and ends coincident with the beginning and ending of the 2D static parallel stripe phase, as a function of concentration.  At a given hole-doping level, superconductivity is entered from a state with static 2D parallel stripe order, as evidenced by quasi-Bragg peaks, and the quasi-Bragg peaks co-exist with superconductivity below T$_C$.  

The superconducting \textit{T}$_c$ vs \textit{x} relationship in Nd-LSCO is structured over this range of concentrations, from 0.05 $\leq$ \textit{x} $\leq$ 0.27, due to the ``$\frac{1}{8}$ anomaly" at \textit{x} = 0.125.  Hence the strongest 2D parallel spin stripe order, in terms of the highest 2D \textit{T}$_N$, appears {\it anti}-correlated with superconductivity.  However, parallel charge stripes onset at temperatures higher still than 2D \textit{T}$_N$, and are intertwined with the parallel spin stripes at and near \textit{x} = 0.125. The charge stripes have been extensively studied in LBCO and to a somewhat lesser extent in Nd-LSCO, but less so still in LSCO - presumably because they are weaker still. The suppression of \textit{T}$_c$ due to the ``$\frac{1}{8}$ anomaly" in LSCO is itself much less pronounced than in either Nd-LSCO or LBCO. One could therefore argue that the conditions for enhanced superconductivity in these single-layer, hole-doped cuprates is the presence of 2D parallel spin stripes, without, or with less-well-developed, parallel charge stripes.

Recent NMR experiments at high magnetic field in LSCO has reported glassy antiferromagnetism existing to higher doping levels than previously believed, up to $\sim$ 0.19 \cite{frachet2019hidden}.  These measurements require a strong magnetic field to quench superconductivity which would otherwise hide the antiferromagnetic NMR signal.  As the enigmatic pseudogap phase in LSCO \cite{LSCOqcp,proust2019remarkable,taillefer2010scattering} itself exists up to \textit{p}$^*$ $\sim$ 0.19, this result implies that strong AF correlations dominate the entire pseudogap phase in LSCO.  

The pseudogap phase is also well studied in the Nd-LSCO system and recent work has shown strong thermodynamic evidence for a pseudogap quantum critical point at \textit{p}$^*$ $\sim$ 0.23 \cite{michon2019thermodynamic}.  Therefore the corresponding argument in Nd-LSCO would have strong static AF correlations existing to even higher hole doping levels, as \textit{p}$^*$ $\sim$ 0.23 in Nd-LSCO.  In fact, this is broadly consistent with the present observation of parallel spin stripes existing in Nd-LSCO over the concentration range 0.05 $\leq$ \textit{x} $\leq$ 0.26.  While \textit{x} = 0.26 is clearly higher than \textit{p}$^*$ $\sim$ 0.23, elastic neutron scattering measures ``static" correlations at much shorter time scales than does NMR (10$^{-10}$ seconds for neutrons compared with 10$^{-6}$ seconds fo NMR).  Consequently spin correlations on time scales falling between neutron ``static" and NMR ``static" would appear static to neutrons and dynamic to NMR.  One may therefore expect to observe static antiferromagnetism with neutron scattering across a broader range of doping than with NMR.

There remains the question of how the pseudogap quantum critical point at \textit{p}$^*$ $\sim$ 0.23 in Nd-LSCO manifests itself in the 2D static parallel spin stripe magnetism that is the subject of this paper.  It is {\it not} through the nature of the quasi-static spin stripe order, which evolves smoothly across the four Nd-LSCO samples studied here, and therefore smoothly across \textit{p}$^*$.  However, this is not the case for the low energy spin fluctuations.   As shown in Fig. \ref{energyscan} and Fig. \ref{ecuts}, the spectral weight of the spin fluctuations for $\hbar\omega$ $<$ 3 meV at \textit{x}=0.19 is qualitatively different from that associated with samples with \textit{x} $>$ \textit{p}$^*$ $\sim$ 0.23: \textit{x} = 0.24 and 0.26.  These low energy inelastic fluctuations are much stronger and more strongly peaked at the IC ordering wavevector at \textit{x} = 0.19 than at \textit{x} = 0.24 and 0.26.  The temperature scale for the crossover or phase transition associated with the pseudogap phase is itself known to be much higher than either 2D \textit{T}$_N$ or superconducting \textit{T}$_C$ in Nd-LSCO \cite{mirelakyle}.  For example this T$^*$ is $\sim$ 100 K at \textit{x}=0.19, and it would seem more reasonable to associate the pseudogap with the relatively high energy scale of these spin fluctuations, rather than the low energy scale of the quasi-static spin stripe magnetism.

\section{Conclusions}

Our new elastic neutron scattering measurements on Nd-LSCO samples with doping levels in the 0.125 $\le$ \textit{x} $\le$ 0.26 range show the presence of AF IC quasi-Bragg peaks at low temperatures, indicating static parallel spin stripe order across this broad range of doping levels.  Combined with earlier neutron scattering measurements on Nd-LSCO and LBCO for \textit{x} $\le$ 0.125, we assemble a composite phase diagram for 2D static parallel spin stripes in single-layer hole-doped cuprates, which covers the concentration range from 0.05 $\le$ \textit{x} $\le$ 0.26.  This range overlaps well with the range of hole-dopings in Nd-LSCO that display superconducting ground states, and show that the superconducting state in Nd-LSCO at any \textit{x} is entered, from above, through a state with 2D parallel spin stripe order. 

Low energy inelastic neutron scattering show that the spin fluctuations captured by this spectral weight are qualitatively different at \textit{x} = 0.19, near optimal doping levels, compared with \textit{x} = 0.24 and 0.26, both of which are above the pseudogap quantum critical point, \textit{p}$^*$ $\sim$ 0.23, and near the end of the superconducting dome.  This spectral weight, $\chi \prime\prime (\textbf{\textit{Q}}= [0.5, 0.64, 0], \hbar\omega = 2 meV)$, displays an ``inelastic order parameter" which grows with decreasing temperature and saturates coincident with superconducting T$_C$ $\sim$ 13.5 K.  The elastic order parameter also shows plateau like behavior at T$_C$, and it can be decomposed to show a phenomenological 2D order parameter that shows the same plateau below T$_C$ as the ``inelastic order parameter". Taken together these new observations at optimum and high doping in Nd-LSCO provide compelling evidence for strong coupling between the superconducting order parameter and that corresponding to parallel spin stripe order and its fluctuations.


\begin{acknowledgments}

We thank Louis Taillefer, Amireza Ataei, Takashi Imai and Graeme Luke for useful and stimulating discussions. We also thank Jianshi Zhou and Zongyao Li for valuable insights into the crystal growth of Nd-LSCO. This work was supported by the Natural Sciences and Engineering Research Council of Canada. We acknowledge the access to neutron infrastructure and supporting technical assistance from the Australian Centre for Neutron Scattering. This research used resources at the Spallation Neutron Source and the High Flux Isotope Reactor, DOE Office of Science User Facilities operated by the Oak Ridge National Laboratory (ORNL). 

\end{acknowledgments}

%

\end{document}